\documentclass[journal]{IEEEtran}
\IEEEoverridecommandlockouts

\usepackage{amsmath,amssymb,amsfonts,amsthm}
\usepackage{graphicx}
\usepackage[dvipsnames,table]{xcolor}
\usepackage{float}
\usepackage{comment}
\graphicspath{{figs/}}
\usepackage{algorithmic}
\usepackage{textcomp}
\usepackage{eurosym}
\usepackage[english]{babel}
\usepackage{enumerate}
\usepackage{siunitx}
\usepackage{pgfplots}
\pgfplotsset{compat=1.18}
\usetikzlibrary{arrows.meta, decorations.markings}
\usepackage[
    hyperindex=true,
    breaklinks=true,
    colorlinks=true,
    bookmarks=false,
    pdftitle={Distributionally Robust Chance-Constrained Scheduling},
    pdfauthor={Quattrociocchi et al.},
    linkcolor=black,
    citecolor=black,
    filecolor=black,
    urlcolor=black,
    plainpages=false,
    pdfpagelabels,
    pagebackref=false
]{hyperref}
\usepackage{svg}
\svgpath{{svg-inkscape/}{figures/}}
\definecolor{lightgray}{gray}{0.9}
\definecolor{ggreen}{rgb}{0,0.5,0}
\definecolor{rred}{rgb}{0.5,0,0}

\theoremstyle{definition}

\newtheorem{dfn}{Definition}

\def\BibTeX{{\rm B\kern-.05em{\sc i\kern-.025em b}\kern-.08em
    T\kern-.1667em\lower.7ex\hbox{E}\kern-.125emX}}

\begin{document}

\title{
%Wasserstein Distributionally Robust Scheduling of Electric Boilers Under Heat Demand Forecast Uncertainty
Distributionally Robust Scheduling of Electrified Heating Under Heat Demand Forecast Uncertainty
\thanks{Email: \{alqua, mantal, pizgo, tomdr\}@dtu.dk. This work is partially supported by the Independent Research Fund Denmark (Danmarks Frie Forskningsfond). Project MAGIC, Grant No. 1051-00003B.}%
}
\author{%
\IEEEauthorblockN{Alessandro Quattrociocchi\textsuperscript{}, Manisha Talukdar\textsuperscript{}, Pere Izquierdo Gómez\textsuperscript{}, and Tomislav Dragi\v{c}evi\'c\textsuperscript{}}\\
\IEEEauthorblockA{\textit{Department of Wind and Energy Systems, Technical University of Denmark (DTU), Kgs. Lyngby 2800, Denmark.}}%
}

\maketitle

\begin{abstract}
Electrified heating systems with thermal storage, such as electric boilers and heat pumps, represent a major source of demand-side flexibility. Under current electricity market designs, balance responsible parties (BRPs) operating such assets are required to submit binding day-ahead electricity consumption schedules, and they typically do it based on forecasts of heat demand and electricity prices. Common scheduling approaches implicitly assume that forecast uncertainty can be well characterized using historical forecast errors. In practice, however, the cumulative effect of uncertainty creates significant exposure to imbalance-price risk when the committed schedule cannot be followed.
To address this, we propose a distributionally robust chance-constrained optimization framework for the day-ahead scheduling of a multi-MW electric boiler using only limited residual forecast samples. We derive a tractable convex reformulation of the problem and calibrate the ambiguity set directly from historical forecast-error data through an a priori tunable risk parameter. Numerical results show that enforcing performance guarantees on the heat-demand balance constraint reduces demand violations by 40\% compared to a deterministic forecast-based scheduler and up to 10\% relative to a nominal chance-constrained model with a fixed error distribution. Further, we show that modeling the real-time rebound cost of demand violations as a second-stage term can reduce the overall daily operating cost by up to 34\% by hedging against highly volatile day-ahead electricity prices.

\end{abstract}
\vspace{0.5em}

\section{Introduction}
The ongoing green electrification is shifting energy production from centralized, fossil fuel-based generation to dispersed renewable electricity sources. Variability in renewable generation increases electricity price volatility, which incentivizes industrial assets to use their flexibility to reduce operating costs, while contributing to grid stability. In this context, electrified heating systems with thermal storage, such as electric boilers and heat pumps, hold a valuable demand-side flexibility potential because they can shift electricity consumption over time while still meeting heat demand~\cite{skalyga2023distributionally, pourmousavi2017real, chen2023study}.

In current electricity market designs, the balance responsible party (BRP) associated with an industrial load such as an electric boiler, a heat pump, or a portfolio of loads and generators submits a day ahead energy schedule to the relevant system operator. The BRP may subsequently rebalance this position through intraday trading or balancing and ancillary services as forecasts and operating conditions change. %The same assets may also provide balancing and ancillary services, in which case system operator activations are settled as balancing energy and reflected via imbalance adjustments. 
Any remaining deviations between the nominated schedule and metered consumption or production are settled at imbalance prices, which can be highly volatile~\cite{talukdar2025real, toubeau2020data}.  

Day-ahead scheduling is therefore highly dependent on the quality of heat-demand and electricity-price forecasts, which has fueled extensive research on short-term heat-demand prediction, ranging from autoregressive methods to multi-step machine-learning models~\cite{runge2023comparison, schneider2025unlocking, potovcnik2021machine}. Although these models often achieve strong average accuracy, forecast uncertainty persists, particularly during demand peaks and atypical operating conditions. Consequently, relying on point forecasts alone can still lead to unacceptably high exposure to unfulfilled load demand and associated imbalance cost.

\begin{comment}
    This stringent energy procurement process ties the efficiency of day-ahead commitments to the accuracy of heat demand forecasts. As a result, short-term prediction has become an expanding research domain, with new methodologies ranging from autoregressive to multi-step-ahead machine learning models~\cite{runge2023comparison, schneider2025unlocking, potovcnik2021machine}. Although these models typically yield high average accuracy, their forecasts are subject to cumulative uncertainty and demand peaks, which are intrinsically challenging to predict ahead of operations and can result in temporary shortages or surpluses.
\end{comment}

Imbalance cost risk mitigation is commonly achieved by purchasing the baseline electricity consumption in the day-ahead market that fully covers expected heat demand. Any residual mismatch between scheduled and realized heat demand, which is not balanced out in intraday or ancillary markets, is then covered by activating backup generation units, such as gas-fired boilers. While effective for ensuring coverage of heat demand, this strategy often increases operating costs and carbon emissions, and reduces the ability to capitalize on periods of low-cost, low-carbon renewable electricity.

An alternative approach is to incorporate forecast uncertainty explicitly during the planning optimization stage. In the literature, these problems are commonly framed as stochastic~\cite{zhang2017chance, liu2025stochastic, liu2015bidding} or robust optimization~\cite{7491274, an2014exploring}.
Robust optimization enforces feasibility for all realizations within a prescribed uncertainty set, which can lead to conservative bid strategies, particularly in volatile markets.
Alternatively, stochastic optimization enforces feasibility through suitable risk measures, such as in chance-constrained problems (CCPs)~\cite{8892469, polisetty2025chance, zhong2021chance}. In CCPs, hard constraints are replaced with probabilistic counterparts, allowing feasibility at a level of $1-\alpha$, where $\alpha$ represents the tolerated risk across a historical set of scenarios.

Exemplifying its application in the literature,~\cite{micheli2023stochastic} proposes a chance-constrained stochastic model predictive controller to lower operational expenses under uncertain thermal and electricity demands using Gaussian processes. Similarly,~\cite{nezhad2022scheduling} develops a robust chance-constrained scheduling model that addresses uncertainties in both generation and consumption, reformulating CCPs as mixed-integer linear programming (MILP).
Other applications in risk management for CCPs include ~\cite{tan2021chance}, which proposes a CVaR-based linearized chance-constrained two-stage optimization, or~\cite{polisetty2025chance}, which studies the dynamic dispatch of electric boilers and the curtailment of renewable energy sources (RES) to address load and renewable uncertainty.

Finally, applications where constraint violations result in direct economic penalties, such as the activation of backup generators, are formulated within a two-stage stochastic programming framework. 
% The feasibility of the balance constraint is ensured through a chance constraint, while the second-stage objective accounts for the expected cost arising from uncertainty in the forecast error realization.
While the feasibility of the balance constraint is ensured through a chance constraint, the second-stage objective accounts for the expected cost arising from uncertainty in the forecast error realization.

However, the scheduling approaches above typically assume that the underlying data-generating distribution can be accurately inferred from historical data, thereby neglecting the risk induced by model misspecification and distribution shifts. In contrast, recently-proposed distributionally robust optimization (DRO) framework provides a principled framework to handle distributional uncertainty~\cite{shafieezadeh2019regularization}. DRO hedges against such uncertainty by optimizing the worst-case expected realization over an ambiguity set of probability distributions defined as a so-called Wasserstein ball of prescribed radius centered at the empirical distribution.

DRO guarantees that the decision remains reliable in cases of misspecification or shifts in the original sample distribution by retaining the distribution of the stochastic variable itself as uncertain. This method, initially applied to address distribution shifts in machine learning, has expanded to a vast number of applications, ranging from machine learning~\cite{kuhn2019wasserstein} to power systems~\cite{poolla2020wasserstein, ordoudis2021energy, hall2024carbon} and data-driven control~\cite{coulson2021distributionally}. 
More recently, DRO applications also gained traction in energy markets and flexibility estimation~\cite{aolaritei2025hedging, skalyga2023distributionally}, to robustify the decision-making processes in the presence of distributional uncertainty.

Building upon this framework, this paper adopts the perspective of a BRP that bears the financial responsibility for the day-ahead scheduling of an electric boiler. In practice, the BRP operates under limited information, having access to only a small set of historical forecast errors, being exposed to seasonal variability, and relying on a load forecasting model provided by a third-party stakeholder. To address this issue, this paper proposes a Wasserstein-ball based distributionally robust chance-constrained (DRCC) formulation for the optimal scheduling of large-scale industrial electric boilers in the day-ahead market, where uncertainty is characterized directly from historical forecast errors via a Wasserstein ambiguity set.

The key methodological and practical contributions are summarized as follows.

\begin{enumerate}

\item We propose a single-stage DRCC formulation for the day-ahead scheduling of an electric boiler, where heat-demand mismatches arising from forecast errors are compensated in real time via a gas boiler. 

\item We explicitly model the real-time rebound cost as a recourse component within the DRCC framework, enabling minimization of the worst-case gas boiler activation cost. 
\end{enumerate}

The proposed framework is applied to the day-ahead scheduling of a multi-MW electric boiler participating in the Danish day-ahead market. Numerical results show that the single-stage DRCC formulation significantly improves out-of-sample performance, reducing the frequency of real-time activations of the backup gas boiler, lowering operational costs, and limiting energy spillage. When extended with a second-stage recourse to explicitly account for gas boiler scheduling costs, the two-stage DRCC formulation enables proactive hedging against highly volatile day-ahead prices, thereby minimizing total expenditure. Overall, the results demonstrate that distributionally robust strategies can systematically unlock the value of thermal flexibility under existing market rules while controlling imbalance risk.

The remainder of the paper is organized as follows. Section II introduces the heating system configuration and the bidding strategy under uncertainty. Section III presents the proposed methodology. Section IV reports the numerical results. Section V concludes the paper with a discussion of the main findings and future research directions.

\section{Problem Description}
We consider a district heating system formed by an industrial electrical boiler with a capacity of $C^{\text{boil.}} = 60$ MWh and a maximum rated power of $p_{\text{max}} = 10$ MW, located in Denmark. 
The boiler under analysis converts electricity with an efficiency of $\eta_{\text{p2h}}$ into heat storage measured in MWh, which is subsequently discharged in response to the heat demand from a storage tank with heat loss factor $\eta_{\text{loss}}$. The heating system is augmented with a gas boiler with a capacity of $C^{\text{gas}} = 1$ MW to counterbalance potential heat shortages during periods of high demand. 

The charging profile of the electric boiler for daily operation is planned one day in advance, and its charging baseline is submitted to the Transmission System Operator (TSO). The boiler's operation is optimized according to heat demand forecasts, which are supplied in the form of receding horizon pointwise trajectories $\mu_t$, based on historical data, exogenous weather data, and seasonality. Heat demand forecast methodology is outside of the scope of this paper, and is the subject of another work. The scheduling framework proposed in this paper adopts the perspective of an aggregator, who is responsible for trading the asset's flexibility in the day-ahead market without direct access to the forecasting model that can be acquired from an external vendor, but only observing heat demand forecasts of the day and historical forecast errors. 

\begin{figure}[H]
    \centering
    \includegraphics[width=\columnwidth]{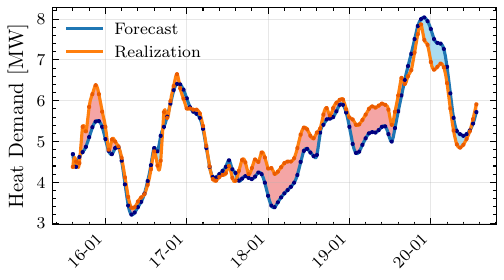}
    \caption{Heat demand forecasts (blue line) versus actual heat demand realizations (red line). The red shaded area indicates upward deviations (i.e., realizations exceeding forecasts), while the blue shaded area indicates downward deviations (i.e., realizations falling below forecasts).}
    \label{fig:prediction_vs_forecast}
\end{figure}

In this setup, the forecast error, $\xi$, is managed by a gas boiler, which can be activated over the day to rebalance the excess of heat demand. As already discussed in the introduction, this strategy is suboptimal, as it can miss low electricity price opportunities, leading to expensive gas activations and in extreme cases with large forecast errors, the heat demand may not be fully met. Intuitively, one could expect that an average accurate forecast, with minimal error, is sufficient to operate the systems reliably. However, in day-ahead scheduling problems, the cumulative forecast errors or unforeseen spikes revealed in real time can lead to suboptimal operation. This is illustrated in Fig.~\ref{fig:prediction_vs_forecast} where residuals larger than the gas boiler's capacity can be observed for extended periods.

\section{Methodology}
The proposed methodology is formulated as a distributionally robust decision-making framework to enhance the boiler's flexibility in the day-ahead market by taking into account the distribution of heat demand residual errors. The electric boiler under analysis operates within a larger heating system, with the role of converting electrical power into thermal energy, which is subsequently discharged based on heat demand forecasts.

Without loss of generality, we assume that the day-ahead trader has limited access to the forecasting model but observes and has access to real-time heat demand realizations in the form of $\mu_t + \xi_t$. The residual, $\xi_t$, is modeled as an additive stochastic error representing the deviation between the realized heat demand and the forecasted value at time $t$. We investigate the design of a decision-making framework with a tunable robustness level on the realization of stochastic forecast error, thus maintaining a bounded risk level in the heat demand feasibility constraints.

\subsection{Problem Formulation}
The decision-making framework is modeled as a CCP to explicitly enforce the feasibility of the constraint at a risk level associated with the realization of the stochastic term $\xi_t$.
Specifically, the model determines the optimal heating schedule, $p^{\text{da}}$, by day-ahead procurement cost subject to physical limits and the requirement that the system remains feasible with a probability of at least $1-\alpha$. The formulation of the CCP is presented in~\eqref{eq:SSCCP}:

\begin{subequations}\label{eq:SSCCP}
\begin{align}
\min_{p^{\text{da}},\, Q^{\text{Stor.}}} \quad 
& \sum_{t=1}^T c_t\, p_t^{\text{da}}
\label{eq:objective} \\[0.2em]
\text{s.t.} \quad 
& \forall\, t=1,\dots,T: \nonumber\\
& \mathbb{P}\!\left( \mu_t + \xi_t \leq H_t \right)
\ge 1-\alpha
\label{eq:chance_constraint} \\
& Q_t^{\text{Stor.}} = (1-\eta_{\text{loss}})\, Q_{t-1}^{\text{Stor.}} + Q_t^{\text{da}} - H_t
\label{eq:delivered_heat_def} \\
& Q_t^{\text{da}} = \eta_{\text{p2h}}\, p_t^{\text{da}}\, \Delta t
\label{eq:energy_power_model} \\
& \underline{p}^{\text{da}} \le p_t^{\text{da}} \le \overline{p}^{\text{da}},
\quad p_t^{\text{da}} \ge 0
\label{eq:power_bounds} \\
& \underline{Q}^{\text{Stor.}} \le Q_t^{\text{Stor.}} \le \overline{Q}^{\text{Stor.}}
\label{eq:storage_bounds} \\
& Q_0^{\text{Stor.}} = Q^{\text{init}}
\end{align}
\end{subequations}

The objective function presented in~\eqref{eq:objective} expresses the cost of purchasing energy on the day-ahead market, where the cost $c_t$ is the day-ahead electricity price in \texteuro/MWh. 
Constraint~\eqref{eq:chance_constraint} imposes a chance energy-balance requirement with confidence level $1-\alpha$. It ensures that, for each time period $t$, the available heat $H_t$, given by the previous storage state including losses, day-ahead heat production, and the stored heat, covers the uncertain heat demand. This is a one-sided chance constraint that limits shortage risk, requiring that, for each $t$, the probability that the uncertain heat demand $\mu_t+\xi_t$ exceeds the available heat $H_t$ is at most $\alpha$. The constraint~\eqref{eq:energy_power_model} describes the relation between the day-ahead power and thermal energy storage scaled by the power-to-heat efficiency term $\eta_{\text{p2h}}$. 

\subsection{Preliminaries}
In the interest of generalization, we refer to the CCP presented in~\eqref{eq:SSCCP} to its compact formulation. Here, the CCP constraint is enforced independently at each time period, resulting in a sequence of single-period distributionally robust chance constraints with scalar uncertainty.
\begin{equation}
    \begin{aligned}
    \min_{x \in \mathbb{R}^n} 
    \quad & c^\top x\\[0.2em]
    \text{s.t.} \quad 
    & \mathbb{P}\left[ g(x, \xi)\leq 0 \right] \geq 1 - \alpha\\
    & \text{Constraints} \eqref{eq:energy_power_model} \text{--} \eqref{eq:storage_bounds}, 
\end{aligned}\label{eq:drcc_problem}
\end{equation}
where $c$ is a real-valued, non-negative cost, $x$ represents the decision variable and $g(x, \xi)$ the stochastic constraint of the optimization problem.  
The CCP as defined in~\eqref{eq:drcc_problem} is, in general, convex only for a restricted set of distributions and constraint functions. Therefore, a well-established convex approximation proposed in the literature relies on the approximation of the CCP using the \textit{Conditional Value-at-Risk} (CVaR)~\cite{nemirovski2007convex}.
This risk measure is generally convex, has a known tractable reformulation, and represents the standard in many risk-averse stochastic problems. A formal definition of the CVaR is presented in Def.~\ref{dfn:cvar_reformulation_def}.

\begin{dfn}[Conditional Value at Risk~{\cite[Th.1]{rockafellar2000optimization}}]\label{dfn:cvar_reformulation_def}
For a random variable $\xi$ under probability measure $\mathbb{Q}$ and a measurable function $g(x,\xi)$, the Conditional Value at Risk (CVaR) at confidence level $1-\alpha$, with $\alpha \in (0,1)$, is defined as:
\begin{align}
\mathrm{CVaR}_{1-\alpha}^{\mathbb{Q}}\!\left(g(x,\xi)\right)
:= \inf_{\tau \in \mathbb{R}}
\left\{ \frac{1}{\alpha}
\mathbb{E}\!\left[(g(x,\xi)-\tau)_+\right]
+ \tau
\right\}.
\end{align}
\end{dfn}

The tractable reformulation presented in Def.~\ref{dfn:cvar_reformulation_def} introduces the auxiliary scalar $\tau$, whereby the risk measure is expressed as the minimum of a linear term and the expected positive deviation of $g(x,\xi)$ above $\tau$.
It is worth noting that the convex reformulation using a $\text{CVaR}$-constrained problem is inherently more conservative, as it prevents the magnitudes $\xi$ beyond the threshold from becoming arbitrarily large by accounting for the expected magnitude of the worst-case $\alpha$ tail violations. 

However, in most of the recent data-driven applications, the true distribution is unknown and rarely entirely described by a single predefined standard distribution.
A common approach in the literature is to exchange the definition of probability distribution with the empirical distribution supported by the training samples to approximate the true unknown distribution, with methods such as Sample Average Approximation (SAA).  These approaches are particularly effective when the sample size $N$ is large enough to accurately represent the true underlying distribution. Whereas, for all the cases with scarce data, uncertain distribution, or corrupted data, the hypothesis of a single data-driven empirical distribution may fail to capture the complexity of the underlying data representation. 
For instance, in DH, the heat demand is influenced by several exogenous factors that contribute to aleatoric uncertainty, such as abrupt temperature drops, changes in user behavior, or public holidays, which are difficult to model in advance in the forecast and are known to cause large imbalances and high rebound costs. 

A suitable solution to mitigate the effect of these distributional shifts is to robustify the CCP by considering a distributionally robust formulation, where the CVaR constraint is enforced for a set of suitable neighbor distributions centered around the empirical distribution defined by the so-called \textit{Wasserstein distance}. 

\begin{dfn}[Wasserstein Distance {\cite[Def.~2]{shafieezadeh2019regularization}}] \label{dfn:wasserstein_distance} ~\\
The type-1 Wasserstein distance between two distributions $\mathbb{Q}_1$ $\mathbb{Q}_2$ supported on $\Xi$ is defined as:
\begin{align}
    d_{W}(\mathbb{Q}_1, \mathbb{Q}_2) = \inf_{\Pi} \Biggl\{\int_{\Xi \times \Xi}||\xi_1 - \xi_2||\:\Pi(d\xi_1, d\xi_2)\Biggl\}\notag,
\end{align}
\end{dfn}
\noindent where $\Pi$ is the joint distribution of $\xi_1$ and $\xi_2$ with marginals $\mathbb{Q}_1$, $\mathbb{Q}_2$.
In words, the Wasserstein distance quantifies the cost of transporting a unit mass from $\xi_1$ to $\xi_2$ under a transportation plan $\Pi$.
This definition shifts the paradigm from considering a single, fixed distribution to a set of candidate distributions within a certain Wasserstein distance from the empirical distribution, known as the \textit{ambiguity set}, as formally defined in~\eqref{dfn:ambiguity_set}.
\begin{align}\label{dfn:ambiguity_set}
\mathcal{B}_\theta(\widehat{\mathbb{P}}_N)
\;:=\;
\left\{
\mathbb{Q} \in \mathcal{M}(\Xi)
\;\middle|\;
d_{W}\!\left(\mathbb{Q},\, \widehat{\mathbb{P}}_N\right) \le \theta
\right\}.
\end{align}
By definition, the ambiguity $\mathcal{B}_\theta(\widehat{\mathbb{P}}_N)$ defines a family of distributions that are within a certain Wasserstein distance $\theta$ from the empirical distribution $\hat{\mathbb{P}}_N$.
Large values of $\theta$ imply higher values of robustness, ensuring that the ambiguity set contains the true unknown distribution with high probability, at the price of more conservative solutions (i.e., higher objective cost). Vice versa, a smaller value of $\theta$ usually results in a smaller objective cost and reduced conservativeness against rare events or larger distributional shifts. Finally, for $\theta=0$, the SAA problem is recovered. 

The following section presents the formulation and its associated tractable reformulation of a distributionally robust chance-constrained problem (DRCCP) by leveraging the definition of Wasserstein distance and ambiguity set introduced above. The DRCCP is applied to an optimal scheduling problem under forecast uncertainty. The formulation presents the problem in the forms of single-stage and two-stage problems. 
In the single-stage problem, the DRCCP is modeled by considering the day-ahead scheduling problem for the electric boiler, while the real-time gas cost activations of the backup boiler are only evaluated in real-time. In this setting, the gas activation costs are not accounted for at scheduling time but only evaluated ex post. 
Conversely, in the two-stage problem, the single-stage problem is extended by incorporating an explicit cost for the violation of the heat demand 
constraint as an additional term in the cost function. This allows a direct estimation of the costs required by the recursive gas activation.

\subsection{Single-Stage Distributionally Robust Chance-Constrained Problem under Additive Uncertainty Representation}\label{subsec:single_stage_drcc} The single-stage DRCCP is formulated in \eqref{eq:ssdrccp}, where the chance constraint is replaced by its CVaR formulation and enforced over a Wasserstein ambiguity set centered at the empirical distribution of historical forecast errors. In this setting, the decision variable $x$ represents the day-ahead power schedule $p^{\text{da}}$, while the random variable $\xi$ denotes a scalar heat-demand forecast error. The uncertainty enters the model additively in the heat balance and is assumed to be independent of the decision variable $x$.

The stochastic constraint function $g(x,\xi)$ is affine in the uncertainty and convex in the decision variable, which allows the distributionally robust CVaR constraint to be reformulated in a tractable form. Specifically, the ambiguity set $\mathcal{B}_\theta(\widehat{\mathbb{P}}_N)$ is defined using the type-1 Wasserstein distance, and the resulting worst-case CVaR remains finite under this affine uncertainty structure, even when the support of $\xi$ is unbounded.

\begin{equation}
\begin{aligned}
 &\min_{x \in \mathbb{R}^n} \quad  c^\top x \\
 \text{s.t.} \quad 
 & \sup_{\mathbb{Q} \in \mathcal{B}_\theta(\widehat{\mathbb{P}}_N)}
\mathrm{CVaR}^{\xi \sim \mathbb{Q}}_{1-\alpha}\!\left(g(x, \xi)\right)
\le 0
\\
& \text{Constraints} \ \eqref{eq:energy_power_model} \text{--} \eqref{eq:power_bounds},
\end{aligned}\label{eq:ssdrccp}
\end{equation}
Under the above modeling choices, Proposition~III.1 in~\cite{poolla2020wasserstein} applies directly, enabling a reformulation of \eqref{eq:ssdrccp} as a tractable linear program via strong duality from optimal transport theory. The resulting formulation, presented in~\eqref{eq:tractable_LP_reformulation}, follows standard derivations in Wasserstein distributionally robust optimization~\cite{shafieezadeh2019regularization, poolla2020wasserstein}.

\begin{equation}\label{eq:tractable_LP_reformulation}
\begin{aligned}
\sup_{\mathbb{Q} \in \mathcal{B}_\theta(\widehat{\mathbb{P}}_N)} &\mathrm{CVaR}_{1-\alpha}^{\xi \sim \mathbb{Q}} \big(g(x, \xi)\big)\le 0\\[0.5em]
&= 
\begin{cases}
&\lambda \theta + \frac{1}{N} \sum_{i=1}^N s_i \leq \tau \alpha \\[0.2em]
&d_k^\top x - f_k + \tau 
    + (e_k - G^\top\gamma_{ik})^{\top} \hat{\xi}_i + \gamma^{\top}_{ik} h \le s_{i}, \\[0.2em]
&\|e_k - G^\top \gamma_{ik}\|_* \le \lambda \\
&\gamma_i \in \mathbb{R}, \ s_i \geq 0,\ \forall i \leq N
\end{cases}
\end{aligned}
\end{equation}

The DRCCP reformulation~\eqref{eq:tractable_LP_reformulation} offers an intuitive interpretation of the distributionally CVaR-robust problem as the combination of \textit{averaging}, \textit{regularization}, and \textit{tightening}, here explained in detail. 

The term $\lambda \theta$ represents the regularization term based on the dual variable $\lambda$ and the Wasserstein distance $\theta$, as a fixed robustness term designed a priori. The higher the value of $\theta$, the higher the robustness level enforced in the problem.  
The term $\frac{1}{N} \sum_{i=1}^N s_i$ captures the average mass of the constraint violations $s_i$. The intuition is the following: whenever a constraint violation occurs, the magnitude of the violation is stored in the variable $s_i$, which is then used as a penalization term. It is evident that when $\theta =0$, the first row resembles the SAA. Moreover, this is a well-known structure in the machine learning field, where the loss function is augmented by a regularization term on the magnitude of the weights.  
Finally, the term $\tau \alpha$ recovers the CVaR level as defined in Def.~\ref{dfn:cvar_reformulation_def}. 
The second row exemplifies the structure of a scalar stochastic constraint, with $d_k^\top x$ the transformation of the decision variable, $f_k$ the right-hand side of the constraint, while the term $(e_k - G^\top\gamma_{ik})^{\top} \hat{\xi}_i + \gamma^{\top}_{ik} h$ includes the transformation of the stochastic variable $\xi$, and its feasibility bounds. In the remainder of the paper, we consider the unbounded uncertainty, enforcing $G=0$ and $h=0$. Therefore, under unbounded uncertainty support ($G=0$, $h=0$), the balance constraint reduces to $d_k^\top x - f_k + \tau + e_k^\top \hat{\xi}_i \le s_i$, where $x$ is the day-ahead power decision variable, $\hat{\xi}_i$ the forecast error in scenario $i$, and $f_k$ the required heat delivery.

\subsection{Two-Stage Distributionally Robust Chance-Constrained Problem under Affine Uncertainty Representation}\label{subsec:two_stage_drcc}
When a constraint violation directly incurs economic costs, the scheduling problem is augmented with a rebalancing second-stage expected cost
in the objective. 
As mentioned in the introduction, the bidding process follows a specific sequence of events: bids are placed in the day-ahead market at D-1, followed by real-time gas activation if demand exceeds storage capacity.

Without gas corrections, the cumulative stochastic realization of the residuals often leads to significant deviations by the end of the day, making it impossible for the backup system to compensate in a single cycle due to the gas boiler's limited capacity. 
To mitigate this cumulative effect and prevent large demand imbalances, the gas-boiler backup is activated in proportion to the forecast error. For instance, if the actual heat demand exceeds the forecast ($\xi >0$), the shortage is addressed by using the gas boiler; conversely, if the demand is lower than forecast ($\xi <0$), the gas boiler remains offline.
Essentially, the gas boiler tracks the real-time realization of the residuals. Consequently, as the gas actuation depends on the outcome of a stochastic variable, the second-stage cost can be modeled as the expected cost $\mathbb{E}[c^\top_2 \cdot y]$, where $y$ represents the real-time activation of the gas boiler in MWh. The decision variable $y$ represents the input actuation of gas activations corresponding to the positive part of the residuals, i.e., the shortage of heat generation relative to the demand. 

In the literature, this second-stage variable ($y$) is traditionally approximated as a linear policy proportional to the uncertainty variable $Y\xi_+$, where $\xi_+$ is the positive part of the residual realization. This transformation can be interpreted as the lack of heat storage faced under the worst-case realization of the uncertainty forecast error in the ambiguity set.
Analogously, the policy $Y\xi_+$ can be interpreted as a second-stage stochastic feedback seen during the scheduling stage. 
Building on a similar formulation introduced for the single-stage DRCCP problem, we define a second-stage cost in~\eqref{eq:two_stages_drcc_problem}.
\begin{equation}\label{eq:two_stages_drcc_problem}
    \begin{aligned}
    \min_{x \in \mathbb{R}^n} 
    \quad & c_1^\top x + \sup_{\mathbb{Q} \in \mathcal{B}_\theta(\widehat{\mathbb{P}}_N)}\mathbb{E}^{\xi \sim \mathbb{Q}}[c_2^\top \cdot Y\xi_+]\\[0.2em]
    \text{s.t.} \quad 
    & \sup_{\mathbb{Q} \in \mathcal{B}_\theta(\widehat{\mathbb{P}}_N)}\mathrm{CVaR}^{\xi \sim \mathbb{Q}}_{1-\alpha}[g(x, \xi)]\le 0\\
    & \text{Constraints} \eqref{eq:energy_power_model} \text{--} \eqref{eq:storage_bounds}, 
\end{aligned}
\end{equation}

Unlike the reformulation presented in~\ref{subsec:single_stage_drcc}, where the reformulation assumed an additive uncertainty in the forecast variable, in the two-stage problem, 
the recourse cost depends affinely on the realization of the forecast error through a linear decision rule. In the single-stage problem, this modeling choice is motivated by the fact that the decision variables do not influence the uncertainty realization of the residuals, which appear as exogenous offsets in the forecast that we aim to robustify against.
Conversely, when the uncertainty realization directly modulates the gas actuation, this dependence is modeled
as a non-additive term in the decision variable. 
This affine dependence between the decision variable $Y\xi$ is often addressed as a game between nature and the decision-maker. In this game, the worst outcome of uncertainty realization is consequent to the decision, with the intention of inflicting the worst possible outcome.

The reformulation of the proposed two-stage distributionally robust chance-constrained problem follows the same optimal-transport-based duality mechanism used in the single-stage case, as presented in~\eqref{eq:tractable_LP_reformulation}. The key distinction lies in the reformulation of the second-stage recourse cost, which depends affinely on the realization of the forecast error and therefore leads to decision-dependent uncertainty. Tractable reformulations of both the worst-case expected recourse cost and the associated CVaR constraint under Wasserstein ambiguity sets are well established in the literature and rely on strong duality results from optimal transport theory. In particular, the developments in~\cite{shafieezadeh2019regularization} and~\cite[Proposition~V.1]{hota2019data} apply directly to the present setting.
The explicit dual reformulations are not reproduced here; instead, the resulting tractable formulation is directly implemented in the numerical experiments. The derivation follows standard two-stage distributionally robust optimization and is consistent with existing formulations in the energy systems literature; the reader is referred to~\cite{ordoudis2021energy} for a detailed derivation of a related two-stage setting.

\section{Results}
We assess the outcome of the distributionally robust single-stage and two-stage chance-constrained models discussed in the previous section with the quality of their out-of-sample performance. In these experiments, we solve the day-ahead boiler schedule generation using the optimization models, followed by the real-time simulation of the actual day of operation for the months spanning from January to March, 2025. The 1 MW gas boiler serves as backup to cover any additional heat demand realized during real-time operations, at a fixed cost of 50 \euro/MWh. We consider a spillage cost of 100 \euro/MWh for overheating the storage tank, which can occur if too much power is scheduled or actual heat demand is lower than expected.

For all the simulations, the CVaR confidence level $\alpha$ is set to 90\%. In this case study, the empirical training samples $\hat{\xi}$ represent the hourly residuals of forecast errors, with G = 0 and h = 0, to indicate that the random variable $\xi$ is unbounded.
The residual error observations are randomly sampled from the historical dataset for a given number of scenarios $N=100$.
In the numerical simulation, to avoid using residual errors from the simulation day in the uncertainty sample set, we exclude all historical observations from the same calendar week as the simulation day. To compare the effectiveness of the distributionally robust approach, all simulations are also carried out with the deterministic scheduling model and the sample average approximation (SAA) model where $\theta = 0$. 

\subsection{Single-stage DRCCP Results}
In the first experiment, we solve the day-ahead scheduling problem using the model formulation of \ref{subsec:single_stage_drcc} which optimizes only the first-stage electricity cost.
The scheduling cost represents the electricity cost, while the total operation cost also includes the a posteriori cost of gas activations and spillage cost on day of operation.
In instances where the actual heat demand cannot be met by either the electric or gas boilers, the average hourly unfulfilled heat (in MW) is considered as a demand violation. 

\begin{figure}[htb]
    \centering
    \includegraphics[width=\columnwidth]{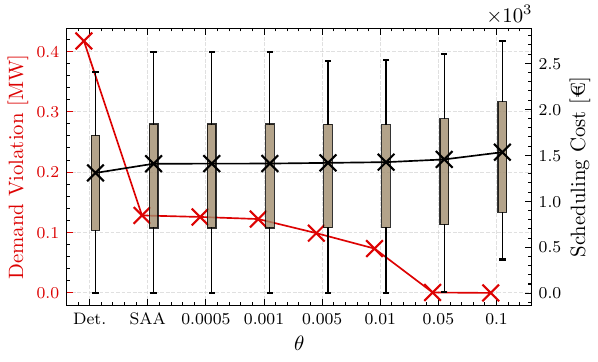}
    \caption{Illustration of the scheduling cost (right) vs. the real-time average violation of the heat demand balance constraint (left axis). The figure shows that increasing values of robustness, depicted by $\theta$ lead to more robust operations, with fewer violations and higher costs.}
    \label{fig:cost_vs_violations}
\end{figure}

Fig.~\ref{fig:cost_vs_violations} visualizes the trade-offs between the out-of-sample cost of electricity procurement and the demand violation between the deterministic, SAA, and single-stage DRCC models for varying Wasserstein radius $\theta$. 

The deterministic model yields the lowest electricity cost; however, because it relies solely on the heat forecast and ignores uncertainty, there are large violations when the actual heat demand cannot be met by the electric boiler's storage or the backup gas boiler. With the SAA approach, the hourly violations decrease by about 0.28 MW relative to the deterministic case, albeit at a higher cost. 
As we increase the Wasserstein radius $\theta$, the DRCC model considers progressively larger distributional shifts from the empirical training samples. This increases the likelihood that the ambiguity set contains the true heat demand distribution of the day. 
This enables the model to make decisions better aligned with the true distribution, minimizing violations at the cost of reserving more power in the day-ahead schedule and thus increasing costs.
% This effectively results in more conservative measurements to handle the stochastic realization of heat demand by reserving more power in the day-ahead schedule, thereby increasing costs.
For example, at $\theta=0.05$ the proposed DRCC policy reduces the average unmet heat demand by approximately 40\% compared to the deterministic scheduling strategy.

\begin{figure}[htbp]
    \centering
    \includegraphics[width=\columnwidth]{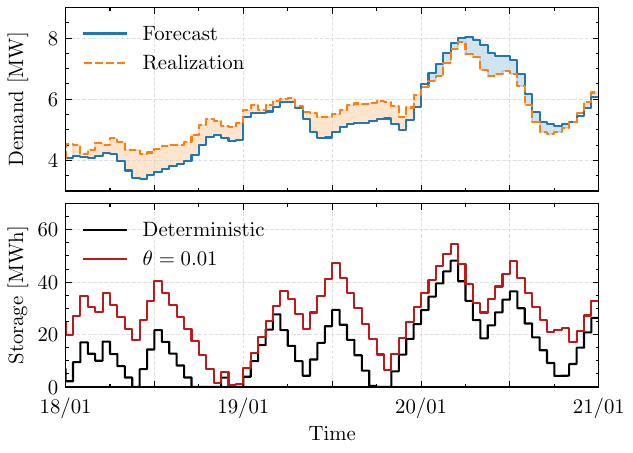}
    \caption{Boiler state of energy comparison under deterministic and DRCCP with $\theta=0.01$ }
    \label{fig:energy_vs_power}
\end{figure}

Fig.~\ref{fig:energy_vs_power} shows a snapshot of the heat demand and the state of energy in the boiler's storage tank from the simulation with the deterministic and single-stage DRCC schedule at $\theta = 0.01$. The top panel shows the deviations of the realized heat demand with respect to the day-ahead forecast. As the deterministic model does not account for any additional heat demand, the energy storage is depleted, particularly towards the end of the 24-hour schedule horizon each day. Most violations, therefore, occur in the final hours of the day due to the cumulative effect of residuals. In the DRCC case, the model allocates more power in the day-ahead market, so the energy storage is sufficient to supply the realized heat demand on the day.

Finally, to prove the effectiveness of the DRCC policy to recover the true unknown distribution under limited error samples, the above simulation is evaluated with a different number of training samples $\hat{\xi}$ from $N \in \{5, 10, 50, 100\}$. In this simulation, the scheduling problem features two degrees of freedom. The first one is represented by the number of training samples $N$, while the second is represented by the Wasserstein radius $\theta$. The quality of scheduling is measured by the out-of-sample operational cost of electricity procurement, backup gas activation and spillage cost calculated in real time.

\begin{figure}[ht!]
    \centering
    \includegraphics[width=\columnwidth]{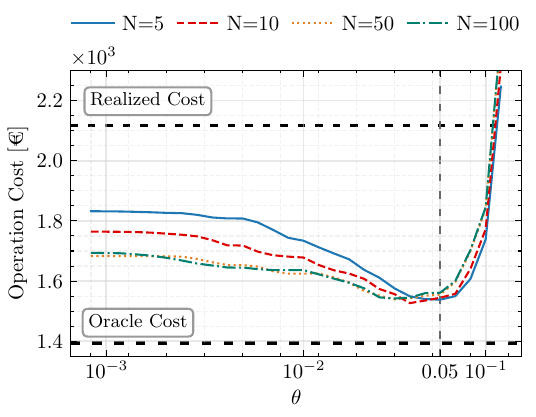}
    \caption{Total out-of-sample operation cost as a function of the Wasserstein radius $\theta$ for different numbers of training samples $N \in \{5, 10, 50, 100\}$. The bottom dashed line represents the oracle cost under perfect forecast information, while the top dashed line denotes the realized cost obtained without accounting for uncertainty. The intermediate curves show the DRCC-based scheduling performance, with a minimum cost around $\theta \simeq 0.05$ that is largely independent of $N$.}
    \label{fig:cost_vs_scenario}
\end{figure}

The results are visualized in Fig.~\ref{fig:cost_vs_scenario}, showing the total operation cost as a function of $\theta$ and $N$.
The bottom dashed line represents the oracle cost, indicating the true cost expected with perfect forecast information. On the opposite end, the top dashed line represents the realized cost, i.e., the total electricity and gas cost from the deterministic model without accounting for the uncertainty. The plots between the oracle and realized cost represent the different out-of-sample approximations of the operation cost function for different values of $\theta$ and $N$.
For smaller values of $\theta$, we observe that the cost drops as we increase the number of samples N. However, this behavior eventually reaches a limit where adding more samples no longer reduces the cost by any significant margin. Even with a large historical dataset of residuals, the SAA approach remains far from achieving the true cost of the problem. 

This limitation is overcome by introducing distribution shift in the empirical training samples by DRCC policy. As $\theta$ increases, the out-of-sample operation costs drops until it reaches a minimum at a critical Wasserstein radius $\theta \simeq 0.05$. For this value, the model achieves an operation cost closest to the true oracle cost, largely independent of the number of training samples considered.

% Conversely, as $\theta$ increases, a larger distribution shift from the empirical training is considered, recovering a candidate of the true heat-demand distribution in the problem. This enables the model to make decisions better aligned with the true distribution of the day, reducing the out-of-sample operation costs until it reaches a minimum at a critical Wasserstein radius $\theta \simeq 0.05$.
% For this value, the model achieves an operation cost closest to the true expected cost, largely independent of the number of training samples considered. 

This demonstrates that by considering a set of suitable distributions in the ambiguity set, the scheduling problem captures a more accurate representation of the true heat demand residuals, even when a small training dataset is available. When applied to real-world case studies with limited historical data, the DRCC formulation is expected to give a better out-of-sample performance than SAA and to hedge against scenarios not well represented in the historical data.

Beyond the critical radius $\theta \simeq 0.05$, the cost increases exponentially due to an increase in violations of the upper limit of the electric boiler's storage tank. The model reserves excessive power in the day-ahead for extreme worst-case scenarios, resulting in overfilling and consequent spillage costs.

\subsection{Performance Comparison of Single-Stage and Two-Stage Decisions}
\begin{figure}[ht]
    \centering
    \includegraphics[width=\columnwidth]{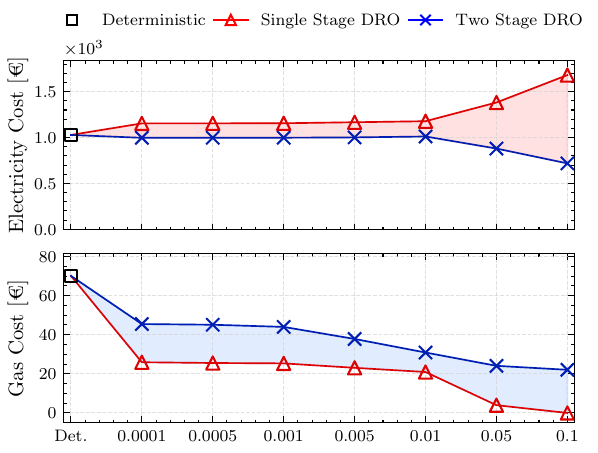}
    \caption{Comparison of average electricity procurement and gas rebound costs for the deterministic, single-stage, and two-stage day-ahead scheduling policies as a function of the Wasserstein radius \(\theta\).}
    \label{fig:cost_between_stages}
\end{figure}

% As a second step in the numerical analysis, 
In this section, we evaluate the deterministic policy against the single-stage and two-stage DRCCP using the model formulation of \ref{subsec:two_stage_drcc}. A cost analysis for varying levels of robustness is illustrated in Fig.~\ref{fig:cost_between_stages}, where the day-ahead electricity scheduling and gas activation costs are averaged across 40 simulation runs.

A prompt observation of the results shows the deterministic model has an average daily electricity cost of €1000. Under this case, the model does not consider uncertainty in the scheduling problem, leading to large demand violations where a portion of the demand is fulfilled at high gas activation cost.
% However, the average gas rebound cost is much higher compared to the DRCC policies. 
Nevertheless, this gas activation is not sufficient to satisfy the true heat demand as shown and discussed in relation to Fig.~\ref{fig:cost_vs_violations}. 

The differences between the single-stage and the two-stage scheduling approaches are evident from the cost profile as a function of the Wasserstein radius $\theta$.
Intuitively, as the single-stage objective is blind to the gas activation costs, it minimizes the average violation by enforcing greater robustness on the electricity scheduling stage. Consequently, this increases electricity costs while reducing the day-of-operation gas costs. 

The two-stage problem optimizes the model over the gas activation cost to counteract demand violations from day-ahead schedule, proactively accounting for the deviations when needed.
% an internal knowledge of the gas rebound costs, proactively accounting for the deviations when needed. 
This knowledge of recourse cost allows the optimizer to trade off between electricity prices with gas prices.
Here, we observe the model actively allows demand violations in the day-ahead, whenever electricity prices are more volatile and economically inconvenient compared to gas activation. This gives a lower electricity scheduling cost with higher gas costs. These violations are allowed under the bounded probability of the heat demand realization, without changing the robustness level of the heat-demand constraint. 
% , resulting in lower electricity costs but higher gas costs.
% This behavior is further exemplified by the attempt to allow violations whenever day-ahead prices are more volatile and economically inconvenient compared to gas activation. These violations are allowed under the bounded probability of the heat demand realization, without changing the robustness level of the heat-demand constraint. 

\begin{figure}[ht]
    \centering
    \includegraphics[width=\columnwidth]{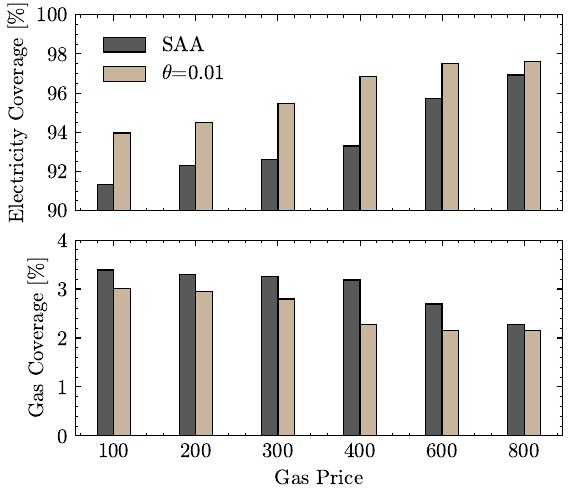}
    \caption{Parametric analysis of the power allocation between the electric and gas boilers as a function of the gas price in the two-stage scheduling problem. As gas prices increase, the optimal allocation progressively shifts from the gas boiler to the electric boiler, with the gas unit being used primarily as a backup resource due to its limited capacity.}
    \label{fig:allocation_vs_gas_cost}
\end{figure}

Finally, the impact of the rebound action cost in the two-stage scheduling problem is evaluated in Fig.~\ref{fig:allocation_vs_gas_cost}.

Here, a parametric analysis of the power allocation between the electric and gas boilers is done for different gas prices.
Results show that during the hours of the day when the gas price is cheaper than the electricity price, the scheduler reserves less day-ahead power, subsequently allowing violations to supply by means of the backup gas boiler during real-time operations. Vice versa, as gas prices increase, the allocation of power shifts from gas to electricity. 

In this case study, given the much smaller gas boiler capacity (1 MW) compared to the electric boiler (10 MW), the power allocation shift represents only a small proportion (1-3\%) of the total heat demand supplied. 
For instance, in the SAA approach, the electric power allocation increases from 91\% to 97\%, while the gas allocation drops from 3.5\% to 2\%. 
Similarly, for $\theta = 0.01$, the electric power allocation, it increases from 94\% to 97.3\%, while the gas allocation drops from 3\% to 2\%. 
These results indicate that whenever the rebound action becomes sufficiently expensive, the allocations tend to converge to similar proportions regardless of the distribution shift applied to the empirical training samples. In other words, for all levels of robustness, the model allocates as much power as possible to the electric boiler and uses the gas boiler as a backup exclusively.

\section{Conclusion}
This paper investigated the day-ahead scheduling of large-scale electrified heating systems under heat-demand uncertainty using a Wasserstein distance based distributionally robust chance-constrained optimization framework. The paper adheres to the perspective of an aggregator that bears the responsibility for bidding a boiler's flexibility in the day-ahead market, relying solely on load forecasts and historical residual errors. The aim of this paper is to present a general data-driven framework that can effectively enforce a level of robustness a priori, enabling asset operators to hedge against rare but impactful demand realizations.

The proposed methodology is applied to a multi-MW electric boiler participating in the Danish day-ahead electricity market. The electric boiler is supplemented by a gas boiler, which compensates for the excess heat demand that exceeds the storage capacity of the electric boiler. A single-stage and two-stage DRCCP are presented. The single-stage problem models day-ahead scheduling by only considering electricity prices, limiting the evaluation of the gas rebound cost to real-time operations. Conversely, the two-stage problem considers the cost of gas rebound activation as part of the scheduling phase.

Numerical results demonstrate that the single-stage DRCC formulation significantly improves out-of-sample performance compared to deterministic and SAA benchmarks. In particular, the proposed approach substantially reduces heat-demand violations and the need for real-time gas boiler activations, at the expense of a moderate increase in day-ahead electricity procurement cost. The analysis further highlights the existence of a critical Wasserstein radius where the closest approximation of the oracle operation cost can be realized, even with a small number of training samples. This illustrates the ability of the DRCC framework to recover the true underlying uncertainty even with limited historical data.

Finally, numerical simulations of the two-stage DRCCP demonstrated that including the rebound cost information during the scheduling phase can help hedge the volatility of the day-ahead price. 
This allows the model to trade off between electricity and gas prices by scheduling demand violations in advance when electricity is more expensive. 
Further, we observe the model optimally shifts the demand proportion between electricity and gas boilers in response to changes in the procurement costs, thus reducing the total operation costs without compromising the feasibility of heat demand.

%\printbibliography
%\vspace{1cm}
\bibliography{references}

@article{poolla2020wasserstein,
  title={Wasserstein distributionally robust look-ahead economic dispatch},
  author={Poolla, Bala Kameshwar and Hota, Ashish R and Bolognani, Saverio and Callaway, Duncan S and Cherukuri, Ashish},
  journal={IEEE Transactions on Power Systems},
  volume={36},
  number={3},
  pages={2010--2022},
  year={2020},
  publisher={IEEE}
}

@article{coulson2021distributionally,
  title={Distributionally robust chance constrained data-enabled predictive control},
  author={Coulson, Jeremy and Lygeros, John and D{\"o}rfler, Florian},
  journal={IEEE Transactions on Automatic Control},
  volume={67},
  number={7},
  pages={3289--3304},
  year={2021},
  publisher={IEEE}
}

@inproceedings{hota2019data,
  title={Data-driven chance constrained optimization under Wasserstein ambiguity sets},
  author={Hota, Ashish R and Cherukuri, Ashish and Lygeros, John},
  booktitle={2019 American Control Conference (ACC)},
  pages={1501--1506},
  year={2019},
  organization={IEEE}
}

@article{ordoudis2021energy,
  title={Energy and reserve dispatch with distributionally robust joint chance constraints},
  author={Ordoudis, Christos and Nguyen, Viet Anh and Kuhn, Daniel and Pinson, Pierre},
  journal={Operations Research Letters},
  volume={49},
  number={3},
  pages={291--299},
  year={2021},
  publisher={Elsevier}
}

@article{shafieezadeh2019regularization,
  title={Regularization via mass transportation},
  author={Shafieezadeh-Abadeh, Soroosh and Kuhn, Daniel and Esfahani, Peyman Mohajerin},
  journal={Journal of Machine Learning Research},
  volume={20},
  number={103},
  pages={1--68},
  year={2019}
}

@article{aolaritei2025hedging,
  title={Hedging against Black Swans in Day-Ahead Energy Markets},
  author={Aolaritei, Liviu and Bangoura, Boubacar and Bolognani, Saverio and Lanzetti, Nicolas and D{\"o}rfler, Florian},
  journal={arXiv preprint arXiv:2510.14328},
  year={2025}
}

@article{rockafellar2000optimization,
  title={Optimization of conditional value-at-risk},
  author={Rockafellar, R Tyrrell and Uryasev, Stanislav and others},
  journal={Journal of risk},
  volume={2},
  pages={21--42},
  year={2000}
}

@article{potovcnik2021machine,
  title={Machine-learning-based multi-step heat demand forecasting in a district heating system},
  author={Poto{\v{c}}nik, Primo{\v{z}} and {\v{S}}kerl, Primo{\v{z}} and Govekar, Edvard},
  journal={Energy and buildings},
  volume={233},
  pages={110673},
  year={2021},
  publisher={Elsevier}
}

@article{runge2023comparison,
  title={A comparison of prediction and forecasting artificial intelligence models to estimate the future energy demand in a district heating system},
  author={Runge, Jason and Saloux, Etienne},
  journal={Energy},
  volume={269},
  pages={126661},
  year={2023},
  publisher={Elsevier}
}

@incollection{kuhn2019wasserstein,
  title={Wasserstein distributionally robust optimization: Theory and applications in machine learning},
  author={Kuhn, Daniel and Esfahani, Peyman Mohajerin and Nguyen, Viet Anh and Shafieezadeh-Abadeh, Soroosh},
  booktitle={Operations research \& management science in the age of analytics},
  pages={130--166},
  year={2019},
  publisher={Informs}
}

@article{skalyga2023distributionally,
  title={Distributionally robust day-ahead combined heat and power plants scheduling with Wasserstein Metric},
  author={Skalyga, Mikhail and Amelin, Mikael and Wu, Qiuwei and S{\"o}der, Lennart},
  journal={Energy},
  volume={269},
  pages={126793},
  year={2023},
  publisher={Elsevier}
}

@article{hall2024carbon,
  title={Carbon-aware computing for data centers with probabilistic performance guarantees},
  author={Hall, Sophie and Micheli, Francesco and Belgioioso, Giuseppe and Radovanovi{\'c}, Ana and D{\"o}rfler, Florian},
  journal={arXiv preprint arXiv:2410.21510},
  year={2024}
}

@article{liu2025stochastic,
  title={Stochastic two-stage multi-objective unit commitment of distributed resource energy systems considering uncertainties and unit failures},
  author={Liu, Jingfan and Zhang, Shijie},
  journal={Reliability Engineering \& System Safety},
  volume={253},
  pages={110520},
  year={2025},
  publisher={Elsevier}
}

@article{zhang2017chance,
  title={Chance-constrained two-stage unit commitment under uncertain load and wind power output using bilinear benders decomposition},
  author={Zhang, Yao and Wang, Jianxue and Zeng, Bo and Hu, Zechun},
  journal={IEEE Transactions on Power Systems},
  volume={32},
  number={5},
  pages={3637--3647},
  year={2017},
  publisher={IEEE}
}

@article{an2014exploring,
  title={Exploring the modeling capacity of two-stage robust optimization: Variants of robust unit commitment model},
  author={An, Yu and Zeng, Bo},
  journal={IEEE transactions on Power Systems},
  volume={30},
  number={1},
  pages={109--122},
  year={2014},
  publisher={IEEE}
}

@article{polisetty2025chance,
  title={Chance Constrained Co-Optimization of Integrated Electrical and District Heating Networks},
  author={Polisetty, Sai Pavan and Nazir, Firdous Ul and Pal, Bikash C},
  journal={IEEE Transactions on Power Systems},
  year={2025},
  publisher={IEEE}
}

@ARTICLE{8892469,
  author={Wang, Bo and Dehghanian, Payman and Zhao, Dongbo},
  journal={IEEE Transactions on Smart Grid}, 
  title={Chance-Constrained Energy Management System for Power Grids With High Proliferation of Renewables and Electric Vehicles}, 
  year={2020},
  volume={11},
  number={3},
  pages={2324-2336},
  doi={10.1109/TSG.2019.2951797}}

@ARTICLE{7491274,
  author={Hu, Wuhua and Wang, Ping and Gooi, Hoay Beng},
  journal={IEEE Transactions on Smart Grid}, 
  title={Toward Optimal Energy Management of Microgrids via Robust Two-Stage Optimization}, 
  year={2018},
  volume={9},
  number={2},
  pages={1161-1174},
  doi={10.1109/TSG.2016.2580575}}

@article{nemirovski2007convex,
  title={Convex approximations of chance constrained programs},
  author={Nemirovski, Arkadi and Shapiro, Alexander},
  journal={SIAM Journal on Optimization},
  volume={17},
  number={4},
  pages={969--996},
  year={2007},
  publisher={SIAM}
}

@article{micheli2023stochastic,
  title={Stochastic MPC for energy hubs using data driven demand forecasting},
  author={Micheli, Francesco and Behrunani, Varsha and Mehr, Jonas and Heer, Philipp and Lygeros, John},
  journal={IFAC-PapersOnLine},
  volume={56},
  number={2},
  pages={11026--11031},
  year={2023},
  publisher={Elsevier}
}

@article{nezhad2022scheduling,
  title={Scheduling of energy hub resources using robust chance-constrained optimization},
  author={Nezhad, Ali Esmaeel and Nardelli, Pedro HJ and Sahoo, Subham and Ghanavati, Farideh},
  journal={IEEE Access},
  volume={10},
  pages={129738--129753},
  year={2022},
  publisher={IEEE}
}

@article{tan2021chance,
  title={Chance-constrained energy and multi-type reserves scheduling exploiting flexibility from combined power and heat units and heat pumps},
  author={Tan, Jin and Wu, Qiuwei and Zhang, Menglin and Wei, Wei and Liu, Feng and Pan, Bo},
  journal={Energy},
  volume={233},
  pages={121176},
  year={2021},
  publisher={Elsevier}
}

@inproceedings{talukdar2025real,
  title={A Real-Time Optimization Framework For Trading Boiler Flexibility on Secondary Reserve Markets},
  author={Talukdar, Manisha and Quattrociocchi, Alessandro and Eliard, Laurena Berangere and Dragicevic, Tomislav},
  booktitle={51st Annual Conference of the IEEE Industrial Electronics Society},
  year={2025},
  organization={IEEE}
}

@article{schneider2025unlocking,
  title={Unlocking Flexibility from Wastewater Treatment Plants Using Machine Learning},
  author={Schneider, Jakob and G{\'o}mez, Pere Izquierdo and Dragicevic, Tomislav},
  journal={IEEE Transactions on Smart Grid},
  year={2025},
  publisher={Institute of Electrical and Electronics Engineers Inc.}
}

@article{pourmousavi2017real,
  title={Real-time demand response through aggregate electric water heaters for load shifting and balancing wind generation},
  author={Pourmousavi, S Ali and Patrick, Stasha N and Nehrir, M Hashem},
  journal={IEEE Transactions on Smart Grid},
  volume={5},
  number={2},
  pages={769--778},
  year={2017},
  publisher={IEEE}
}

@article{chen2023study,
  title={Study of combined heat and power plant integration with thermal energy storage for operational flexibility},
  author={Chen, Chengxu and Ge, Zhihua and Zhang, Youjun},
  journal={Applied Thermal Engineering},
  volume={219},
  pages={119537},
  year={2023},
  publisher={Elsevier}
}

@article{toubeau2020data,
  title={Data-driven scheduling of energy storage in day-ahead energy and reserve markets with probabilistic guarantees on real-time delivery},
  author={Toubeau, Jean-Fran{\c{c}}ois and Bottieau, J{\'e}r{\'e}mie and De Gr{\`e}ve, Zacharie and Vall{\'e}e, Fran{\c{c}}ois and Bruninx, Kenneth},
  journal={IEEE Transactions on Power Systems},
  volume={36},
  number={4},
  pages={2815--2828},
  year={2020},
  publisher={IEEE}
}

@article{zhong2021chance,
  title={Chance constrained scheduling and pricing for multi-service battery energy storage},
  author={Zhong, Weifeng and Xie, Kan and Liu, Yi and Xie, Shengli and Xie, Lihua},
  journal={IEEE Transactions on Smart Grid},
  volume={12},
  number={6},
  pages={5030--5042},
  year={2021},
  publisher={IEEE}
}

@article{liu2015bidding,
  title={Bidding strategy for microgrid in day-ahead market based on hybrid stochastic/robust optimization},
  author={Liu, Guodong and Xu, Yan and Tomsovic, Kevin},
  journal={IEEE Transactions on Smart Grid},
  volume={7},
  number={1},
  pages={227--237},
  year={2015},
  publisher={IEEE}
}
%\section*{Appendix}

\end{document}